\documentclass[10pt,conference]{IEEEtran}

\usepackage[utf8]{inputenc}
\usepackage[T1]{fontenc}
\usepackage{amsmath,amssymb,amsfonts}
\usepackage{bm}
\usepackage{graphicx}
\graphicspath{{figures/}}%
\usepackage{mathtools}

\usepackage[ruled,lined,procnumbered]{algorithm2e}     % algorithm pseudocode
\SetAlgoProcName{Subroutine}{Subroutine}
\SetAlFnt{\small}
\SetAlCapFnt{\small}
\SetAlCapNameFnt{\small}

\SetCommentSty{smallcomment}
\usepackage{float}
\usepackage{caption}

\usepackage[table]{xcolor}
\usepackage{textcomp}
\usepackage{hyperref}
\usepackage[nameinlink,noabbrev]{cleveref}

\usepackage{url}
\usepackage{cite}
\usepackage{booktabs}         % professional tables
\usepackage{subcaption}%  % subfigure-environment

\newcommand*{\C}{\mathbb{C}}							% Complex numbers

\newcommand*{\hil}{\mathcal{H}} 					% Hilbert space

\newcommand*{\id}{I} 											% Identity operator
\DeclareMathOperator{\tr}{tr}
\newcommand{\adj}[1]{{#1}^\dag}%								% Adjoint
\newcommand*{\bounded}[1]{\mathcal{B}(#1)}    % Bounded linear operators

\DeclarePairedDelimiter\abs{\lvert}{\rvert}%					% Absolute value

\hypersetup{
  pdfauthor={},
  pdftitle={},
  pdfsubject={IEEE Quantum Week 2026 (QCE26)},
  pdfkeywords={quantum computing},
  colorlinks=true,
  linkcolor=black,
  citecolor=black,
  urlcolor=blue,
}

\title{Orkan: Cache-friendly simulation of quantum operations on hermitian operators}

\author{
  \IEEEauthorblockN{Timo Ziegler}
  \IEEEauthorblockA{
    Institut f\"ur Theoretische Physik \\
    Leibniz Universit\"at Hannover, Germany \\
    timo.ziegler@itp.uni-hannover.de
  }
}

\begin{document}
% Named data (typeset in sans-serif by algorithm2e)
\SetKwData{Tzz}{T\textsubscript{00}}% % Pointer to the first element of the tile (ti=0,tj=0)
\SetKwData{Tzo}{T\textsubscript{01}}% % Pointer to the first element of the tile (ti=0,tj=1)
\SetKwData{Toz}{T\textsubscript{10}}% % Pointer to the first element of the tile (ti=1,tj=0)
\SetKwData{Too}{T\textsubscript{11}}% % Pointer to the first element of the tile (ti=1,tj=1)
\SetKwData{H}{h}

\maketitle

% --------------------------------------------------------------------------
%  ABSTRACT
% --------------------------------------------------------------------------
\begin{abstract}
  Classical simulation of quantum operations is essential for algorithm design, noise characterisation, and benchmarking of quantum hardware.
  The most general physically realisable operation can be described by a positive linear map acting on a hermitian operator, representing either a density matrix or an observable.
  Established simulators vectorise the density matrix on an \( n \)-qubit Hilbert space and reuse state-vector kernels, storing all \( 2^{2n} \) elements and forgoing the benefits of hermitian symmetry.
  In this work, I introduce \emph{Orkan}(\url{https://github.com/Timo59/orkan}), a simulation library that uses a tiled memory layout storing only the lower triangle of the hermitian matrix at tile granularity, roughly halving both the memory footprint and the wall time to simulate the evolution of quantum states under generic quantum operations.
  The implementation treats any hermitian operator uniformly and is agnostic to whether the Schr\"{o}dinger or Heisenberg picture is used.
  Dedicated \( k \)-local conjugation algorithms update all entries of the hermitian matrix in a single pass.
  Benchmarks against Qiskit Aer, QuEST, and Qulacs show consistent wall-clock speedups of \( 2 \)--\( 4{\times} \) partly attributable to the reduced memory footprint.
\end{abstract}

% --------------------------------------------------------------------------
%  KEYWORDS
% --------------------------------------------------------------------------

% ==========================================================================
\section{Introduction}
\label{sec:Introduction}
% ==========================================================================

Classical simulation of quantum computers is an indispensable tool for quantum algorithm design, verification, and benchmarking.
A simulator can apply any completely positive map---unitary gates, noisy channels, and mid-circuit measurements alike---to the system descriptor, providing exact predictions unconstrained by hardware availability.
In practice, every physical gate on real hardware, be it a calibrated microwave pulse or a laser-driven transition, is itself a completely positive map whose unitary target is only approached in the zero-error limit.
This is particularly valuable for algorithms that rely on more general quantum operations (e.g., mid-circuit measurement-induced LCU channels) as computational primitives~\cite{Binkowski2025_FromBarrenPlateausThroughFertileValleysConicExtensionsOfParameterisedQuantumCircuits, Binkowski2024_OneForAllUniversalQuantumConicProgrammingFrameworkForHardConstrainedCombinatorialOptimizationProblems} or simulating small quantum error correcting codes under realistic noise~\cite{Fowler2012_SurfaceCodes}.
Classical simulation at the level of quantum operations is therefore the natural setting to model any quantum device.

Several paradigms exist for classical simulation of quantum circuits.
State-vector simulators can only approximate the effect of a quantum operation by an ensemble of stochastic pure-state trajectories~\cite{Dalibard1992_WaveFunctionApproachToDissipativeProcessesInQuantumOptics, Molmer1993_MonteCarloWaveFunctionMethodInQuantumOptcis}, trading exactness for a reduced memory footprint.
Stabiliser simulators exploit the Gottesman--Knill theorem for efficient simulation of Clifford circuits~\cite{Gottesman1998_TheHeisenbergRepresentationOfQuantumComputers, Aaronson2004_ImprovedSimulationOfStabilizerCircuits}, decision-diagram simulators compress the quantum state by exploiting redundancy in the amplitude structure~\cite{Viamontes2003_ImprovingGateLevelSimulationOfQuantumCircuits, Zulehner2018_AdvancedSimulationOfQuantumComputations}, and tensor-network methods assume low-entangled states~\cite{Vidal2003_EfficientClassicalSimulationOfSlightlyEntangledQuantumComputations, Brennan2021_TensorNetworkCircuitSimulationAtExascale}.
I refer to~\cite{Young2023_SimulatingQuantumComputationsOnClassicalMachines} for a recent review.
Each of these methods trades generality for efficiency; in particular, generic quantum operations break the structural assumptions that make them tractable, be it sparsity, stabiliser structure, amplitude redundancy, or low entanglement.

\begin{figure}
  \centering
  \includegraphics[width=\columnwidth]{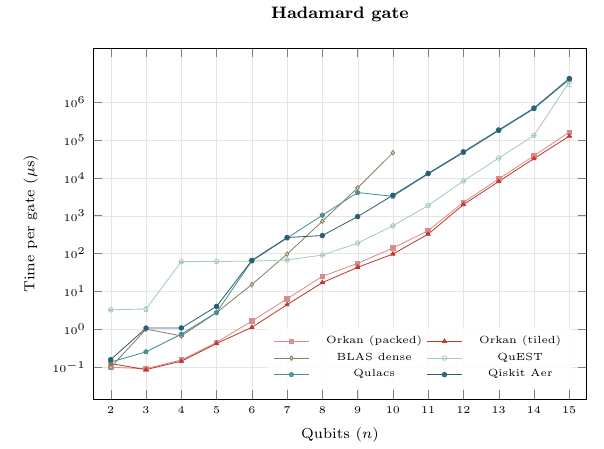}
  \caption{
    Time per execution of the Hadamard gate averaged over all possible qubit positions (\emph{less is better}, cf. \hyperref[sec:Appendix]{Appendix~\ref{sec:AppendixA}}).
  }
  \label{fig:BenchHadamard-a}
\end{figure}

The direct evolution of the density matrix remains the only deterministic, exact method that makes no structural assumptions for predicting the statistics of real quantum circuits.
Quantum computation is operationally a statistical experiment: every prediction is an expectation value \( \tr(\rho\,O) \), and any completely positive map in the middle of the circuit can be absorbed into the state (\( \rho \mapsto \Phi(\rho) \), the \emph{Schr\"{o}dinger picture}) or into the observable (\( O \mapsto \Phi^{\ast}(O) \), the \emph{Heisenberg picture}) with identical predictions.
The fundamental object of a simulator is therefore a hermitian operator evolving under a completely positive map.
Yet in practice, established state-vector simulators~\cite{Jones2019_QuESTandHighPerformanceSimulationOfQuantumComputers, Suzuki2021_Qulacs} vectorise the density matrix into a \( 2n \)-qubit state and reuse state-vector kernels, inheriting their optimisations, but forgoing any exploitation of hermitian symmetry.

This work introduces \emph{Orkan}, a simulation library for qubit-based quantum processors utilising a tiled memory layout for hermitian matrices.
It stores only the lower triangle of the hermitian matrix at tile granularity, roughly halving the memory footprint and is completely agnostic to the choice of Schrödinger or Heisenberg picture.
Combined with dedicated \( k \)-local conjugation algorithms---generic ones driven by a precomputed transfer matrix and specialised paths for native gates---that update each invariant subspace in a single pass, this approach achieves consistent wall-clock speedups of \( 2 \)--\( 4{\times} \) over density-matrix backends (cf. \Cref{fig:BenchHadamard-a}).

The remainder of the paper is organised as follows.
\Cref{sec:Background} establishes the \( k \)-local channel formalism and the gather--transform--scatter kernel.
\Cref{sec:Method} introduces the tiled format and derives the intra-tile and cross-tile update algorithms.
\Cref{sec:Results} benchmarks the implementation against Qiskit Aer, QuEST and Qulacs.
\Cref{sec:Conclusion} discusses the results and gives an outlook.

% ==========================================================================
\section{Background}
\label{sec:Background}
% ==========================================================================

The most general physically realisable operation on a quantum system, described by some Hilbert space \( \hil \),  is a completely positive, trace non-increasing (CP-TNI) linear map \( \Phi: \bounded{\hil} \to \bounded{\hil} \).
A quantum operation \( \Phi \) on an \( n \)-qubit quantum system admits a Kraus representation
\begin{equation}\label{eq:KrausRep}
  \Phi(H) = \sum_{\alpha=1}^{r} K_{\alpha} H K_{\alpha}^{\dag},
\end{equation}
where \( r \) is the Kraus rank and \( H \) is a \( N \times N \) complex matrix with \( N = 2^{n} \).
States and observables are both elements of \( \bounded{\hil} \) and the map acts on either; the Schr\"{o}dinger and Heisenberg pictures are related by duality with respect to the trace inner product \( \langle A, B \rangle = \tr(\adj{A} B) \).

  \begin{procedure}[t]
      \DontPrintSemicolon
      \SetKwInOut{Input}{Input}%	% Inputs
      \SetKwInOut{Output}{Output}%	% Outputs
      \SetKwData{Global}{global}%	% Index on all qubits
      \SetKwData{Pos}{pos}%	% loop variable
      \SetKwData{Right}{right}%	% Right bit mask
      \SetKwData{Left}{left}%	% Left bit mask
      \Input{Subspace label \( s \); local index \( a \); sorted active qubit positions \( A \)}
      \Output{Global index (the bits of \( a \) inserted into \( s \) at positions \( A \))}
      \BlankLine
      \( \Global \gets s \)\;
      \( \ell \gets 0 \)\;
      \For{\Pos in \( A \)} {
        \( \Right \gets ((1 \ll \Pos) - 1) \mathbin{\&} \Global \)\;
        \( \Left \gets (\Global \gg \Pos) \ll 1 \)\;
        \( \Left \mathrel{|}= (a \gg \ell) \mathbin{\&} 1 \)\;
        \( \Left \mathrel{\ll}= \Pos \)\;

        \BlankLine
        \( \Global \gets \Left \mathbin{|} \Right \)\;
        \( \ell \gets \ell + 1 \)\;
      }
      \Return \Global \;
    \caption{insertBits($s$,\,$a$,\,$A$)\label{sub:InsertBits}}
  \end{procedure}

  \subsection{\texorpdfstring{\( k \)}{k}-local quantum operations}

  When the operation \( \Phi \) is \( k \)-local, i.e., it acts non-trivially on a subset \( A \subseteq [0, n-1] \) of \( k \) qubits, every Kraus representation is unitarily equivalent to one whose Kraus operators factorise as \( K_{\alpha} = \id \otimes L_{\alpha} \) with \( L_{\alpha} \in \mathbb{C}^{2^{k} \times 2^{k}} \).
  This is due to the tensor-product structure of the corresponding \emph{Choi} operator~\cite{Watrous_QuantumInformationTheory}.
  This induces a decomposition of the \( n \)-qubit Hilbert space \( (\C^{2})^{\otimes n} \) into \( 2^{n-k} \) mutually orthogonal invariant subspaces \( \mathcal{K}_{s} = \mathrm{span}\{|s\rangle\} \otimes (\C^{2})^{\otimes k} \), so that \eqref{eq:KrausRep} reduces to independent \( 2^{k} \times 2^{k} \) block updates
  \begin{equation}\label{eq:BlockUpdate}
    \bm{h}_{s,s'} \;\longmapsto\; \sum_{\alpha} L_{\alpha}\, \bm{h}_{s,s'}\, L_{\alpha}^{\dag}, \qquad s, s' = 0, \dots, 2^{n-k}-1.
  \end{equation}
  Iterating over all \( 2^{2(n-k)} \) submatrices \( h_{\bm{s}, \bm{s}'} \) yields \( \mathcal{O}(r\,N^{2}\,2^{k}) \) arithmetic operations and \( \mathcal{O}(r\,N^{2}) \) memory writes as opposed to \( \mathcal{O}(r\,N^{3} ) \) operations for the naive dense matrix multiplication with accumulation.
  Since \( r \le 2^{2k} \)~\cite{Watrous_QuantumInformationTheory}, the computational cost is bounded by \( \mathcal{O}(N^{2}\,2^{3k}) \).

  The subspace label \( s \) identifies those elements that are coupled by the \( k \)-local operation.
  In a fixed computational-basis representation, however, the bits encoding the active qubits are scattered throughout the binary index; elements belonging to the same invariant block are therefore non-contiguous in memory.
  Rather than permuting the \( N \times N \) matrix \( H \), which induces an \( \mathcal{O}(N^{2}) \) overhead per gate, global indices are reconstructed on the fly by inserting the \( k \) active-qubit bits into the subspace label using \( \mathcal{O}(k) \) bitwise operations per index~(\hyperref[sub:InsertBits]{Subroutine~\ref*{sub:InsertBits}}).

  \begin{algorithm}[t]
    \DontPrintSemicolon
    \KwData{
      Hermitian \( N \times N \)-matrix \( H \);
      complex \( 2^{2k} \times 2^{2k} \) transfer matrix \( S \);
      ordered set of active qubits \( A \)
    }
    \KwResult{Hermitian \( N \times N \)-matrix \( H' = \Phi(H) \), where \( \Phi \) is the \( k \)-local quantum operation with transfer matrix \( S \).}
    
    \BlankLine
    \tcc{Iterate block rows}
    \For{\( s \gets 0 \) \KwTo \( 2^{n-k} - 1 \)} {

      \BlankLine
      \tcc{Iterate block columns}
      \For{\( s^{\prime} \gets 0 \) \KwTo \( 2^{n-k} - 1 \)} {

        \BlankLine
        \tcc{Gather block elements}
        \For{\( a \gets 0 \) \KwTo \( 2^{k} - 1 \)} {
          \( i \gets\; \)insertBits(\( s \), \( a \), \( A \))\;
          \For{ \( a^{\prime} \gets 0 \) \KwTo \( 2^{k} - 1 \)} {
            \( j \gets\; \)insertBits(\( s' \), \( a' \), \( A \))\;
            \( v_{a + a' \cdot 2^{k}} \gets h_{i,j}\) \;
          }
        }

        \BlankLine
        \( w \gets S v \)\;

        \BlankLine
        \tcc{Update block elements}
        \For{\( a \gets 0 \) \KwTo \( 2^{k} - 1 \)} {
          \( i \gets\; \)insertBits(\( s \), \( a \), \( A \))\;
          \For{\( a^{\prime} \gets 0 \) \KwTo \( 2^{k} - 1 \)} {
            \( j \gets\; \)insertBits(\( s' \), \( a' \), \( A \))\;
            \( h_{i,j} = w_{a + a' \cdot 2^{k}} \)\;
          }
        }
      }
    }
    \caption{Update of the hermitian matrix \( H \mapsto \Phi(H) \) according to a quantum operation \( \Phi \) that acts locally on qubit indices in the subset \( A \subseteq [0, n-1] \).}
    \label{alg:k-localKraus}
  \end{algorithm}

  Additionally, since
  \begin{equation}\label{eq:KrausMap}
    [\bm{h}_{s, s'}]_{ij} \mapsto \sum_{\alpha} \sum_{\mu, \nu} \, [L_{\alpha}]_{i \mu} \, [\bm{h}_{s, s'}]_{\mu \nu} \, \overline{[L_{\alpha}]_{j \nu}},
  \end{equation}
  each block update can be expressed by the matrix vector multiplication \( w = Sv \), where \( S = \sum_{\alpha} \overline{L_{\alpha}} \otimes L_{\alpha} \) is the complex \( 2^{2k} \times 2^{2k} \) \emph{transfer matrix} (or Liouville representation) of the quantum operation and \( v \equiv \mathrm{vec}(\bm{h}_{s, s'}) \) is the vectorisation (stacking columns) of the submatrix \( \bm{h}_{s, s'} \).
  This enables a \emph{gather--transform--scatter} kernel that collects each \( 2^{k} \times 2^{k} \) block into fast memory, applies the Kraus sum via a precomputed transfer matrix \( S \), and writes the result back (\Cref{alg:k-localKraus}).
  It yields \( \mathcal{O}(2^{4k}) \) arithmetic operations per block, which is strictly better than \( \mathcal{O}(r\,2^{3k}) \) once \( r > 2^{k} \) and the precomputation cost of \( S \) is amortised over many blocks.
  Nevertheless, for unitary transformations (\( r = 1 \)), the direct block update~\eqref{eq:BlockUpdate} is asymptotically competitive and retains direct access to \( L \), enabling per-gate optimisation (e.g., zero-arithmetic permutations) that the transfer matrix would obscure.
  This motivates the dedicated treatment of native gates.

\subsection{Native local quantum gates}

  For native quantum gates---unitary transformations that constitute the fundamental building blocks of quantum circuits---it is often beneficial to implement dedicated code paths rather than invoking the implementation for generic quantum operations with a precomputed transfer matrix.
  For example, the Pauli-\( X \) gate avoids the gather and scatter step altogether, and the conjugation reduces to pair-wise memory swaps with zero arithmetic operations, i.e., \( h_{i,j} \leftrightarrow h_{i \oplus 2^{a}, j \oplus 2^{a}} \) and \( h_{i \oplus 2^{a}, j} \leftrightarrow h_{i, j \oplus 2^{a}} \).
  Analogously, the controlled Pauli-\( X \), the SWAP and the Toffoli gate are pure permutation matrices, i.e., their entries are all \( 0 \) or \( 1 \), and require no multiplication at all.
  The Pauli-\( Y \) gate, on the other hand, swaps amplitudes just as the Pauli-\( X \) gate and negates the off-diagonal block elements.
  Diagonal phase gates, e.g., Pauli-\( Z \), \( S \), \( T \) or \( RZ \), leave diagonal elements unchanged for \( U H \adj{U} \) and complex multiplications in off-diagonal elements can be implemented via real-valued shuffles, i.e., swaps and negations of real and imaginary parts with potential real multiplications.

\subsection{Prior work on simulating density matrices}

  Established density-matrix simulators, e.g.,~\cite{Jones2019_QuESTandHighPerformanceSimulationOfQuantumComputers, Suzuki2021_Qulacs}, commonly treat the vectorised density matrices \( \mathrm{vec}(\rho) \) as \( 2n \)-qubit state-vectors and apply the generic quantum operation by its transfer matrix as a \( 2k \)-qubit gate acting on qubits \( A \) and \( \{ a_{i} + n\, : \, a_{i} \in A \} \).
  The dedicated code paths for native quantum gates on state-vectors can be reused at the expense of two separate passes \( (\bar{U} \otimes \id)(\id \otimes U)\,\mathrm{vec}(\rho) \), and the associated HPC optimisations carry over.

  Updating the invariant subspaces is highly parallelisable---this insight led to distributing state-vector simulation across multiple processors~\cite{Obenland1998_AParallelQuantumComputerSimulator, DeRaedt2007_MassivelyParallelQuantumComputerSimulator, Imamura2022_mpiQulacs} already in the late \( 90 \)s.
  On a single machine, multithreading~\cite{Nemirovsky2013_MultithreadingArchitectures} allows for the concurrent update of the individual blocks stored to a shared memory involving multiple logical CPUs.
  The performance of the multithreaded state-vector simulation is critically shaped by the memory hierarchy of modern CPUs~\cite{Bottomly2004_UnderstandingCaching} necessitating the optimisation of memory layouts to avoid \emph{cache-misses}~\cite{Monil2020_UnderstandingTheImpactOfMemoryAccessPatternsInIntelProcessors} and \emph{false-sharing}~\cite{Bolosky1993_FalseSharingAndItsEffectonSharedMemoryPerformance}.
  Finally, uniform update rules within each invariant subspace admit \emph{Single-Instruction Multiple-Data (SIMD)} vectorisation~\cite{Larsen2000_ExploitingSuperwordLevelParallelismWithMultimediaInstructionSet, Shin2005_SuperwordLevelParallelismInThePresenceOfControlFlow} using wide vector registers if the loop structures maximise contiguous access patterns.
  For this purpose, complex arithmetic in \eqref{eq:BlockUpdate},\,\eqref{eq:KrausMap} is separated into real and imaginary parts since the interleaved complex-type structure and complex multiply-adds break the simple SIMD patterns.

\medskip
Despite rare exceptions (e.g.~\cite{Li2020_DensityMatrixQuantumCircuitSimulationViaTheBSPMachineOnModernGPUClusters}), the simulation of density matrices, let alone general hermitian operators (\emph{observables}), receives considerably less attention than state-vectors.
This is a remarkable observation, given the fact that the state of a quantum system is only fully determined by a density matrix and its evolution in an experiment is equally described by the evolution of the observable (Schr\"{o}dinger vs. Heisenberg picture).
All strategies store the full \( N^{2} \) elements, forgoing the \( {\sim}50\% \) memory saving that hermiticity would permit.

% ==========================================================================
\section{Method}
\label{sec:Method}
% ==========================================================================

  \begin{figure*}[t]
    \includegraphics[width=0.95\textwidth]{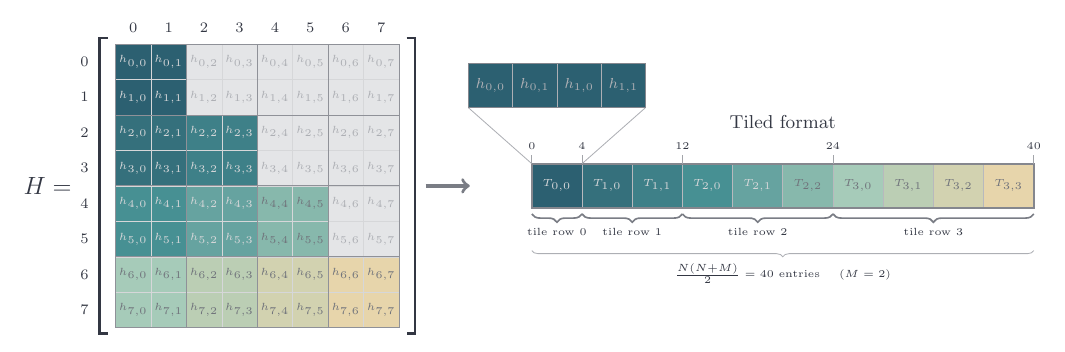}
    \caption{
      The \emph{tiled format} with a tile size of \( M = 2 \).
      The hermitian matrix \( H \) is covered with tiles of edge length \( M \) and only tiles \( T_{t_{i}, t_{j}} \) in the lower-triangle, i.e., with \( t_{i} \ge t_{j} \) are stored to a contiguous array.
      Within the tiles, all elements are stored in row-major format to enhance SIMD vectorisation.
    }
    \label{fig:TiledFormat}
  \end{figure*}

An operator \( H \in \bounded{\hil} \) is called \emph{hermitian} if it satisfies \( H = \adj{H} \).
This implies that any hermitian operator on \( \hil_{n} \) is determined by the \( N (N+1)/2 \) elements in the lower triangle of any matrix representation with respect to an orthonormal basis of \( \hil_{n} \), i.e., by \( h_{ij} \) with \( i \ge j \).
Moreover, since \( \adj{\Phi(H)} = \sum_{\alpha} \adj{(K_{\alpha} H \adj{K_{\alpha}})} = \sum_{\alpha} K_{\alpha} \adj{H} \adj{K_{\alpha}} = \Phi(H) \), any Kraus map preserves hermiticity, so storing only the lower triangle is consistent with the update.
High-performance computing techniques can improve performance beyond the asymptotic regime.
The key design question is therefore how to lay out the hermitian matrix \( H \) in memory to minimise both, the memory footprint and the wall time of the operation.

\subsection{Packed storage and its limitations}
\label{sec:PackedLimitations}

  The straightforward layout--a \emph{packed column-major} array of the \( N(N{+}1)/2 \) lower-triangular elements---introduces three performance pathologies.
  First, since elements in the upper triangle of \( H \) cannot be accessed directly but need to be computed from \( h_{ij} = \overline{h_{ji}} \), \Cref{alg:k-localKraus} is forced to branch in the innermost loop.
  Second, for an operation on qubit \( a \), the local update couples columns at a logical distance of \( 2^{a} \cdot (2(N-j) - 2^{a} - 1) / 2 \) in contiguous memory.
  This access pattern, which is inherent to the packed format, inevitably leads to cache misses.
  Lastly, parallelisation suffers from variable-length column boundaries, i.e., earlier columns carry most of the work because column \( j \) contains \( N-j \) elements.
  Conversely, balancing the load through a \emph{round-robin} assignment, causes false sharing because adjacent columns are stored contiguously.
  The tiled format introduced in the next subsection eliminates per-element branching and false sharing entirely; only the non-contiguous access pattern within intra-tile operations cannot be removed on a fixed memory layout, but its impact is limited to the L1 working set.

\subsection{Tiled format}
\label{sec:TiledFormat}

  In order to improve cache locality, the matrix is covered with square tiles of edge length \( M = 2^{m} \) and only tiles \( T_{t_{i}, t_{j}} \) in the lower triangle of the \( N_{t} \times N_{t} \) (\( N_{t} = N/M \)) tile grid are stored (\Cref{fig:TiledFormat}).
  If \( N < M \), \( H \) is extended by \( M - N \) rows and columns of zeros.
  Tiles are arranged in row-major order in the contiguous array; likewise the elements within each tile are stored in row-major order.%
  \footnote{
    The intra-tile row-major layout is performance-driven, as it enables SIMD-friendly contiguous access within each tile---a prerequisite for auto-vectorisation.
    The row-major arrangement of tiles in the grid itself is arbitrary and could equally well be column-major.
  }
  While off-diagonal tiles contain only lower-triangular elements of \( H \) by construction, diagonal tiles store all \( M^{2} \) elements to eliminate per-element branching.
  Together with the contiguous access, the decomposition of complex arithmetic, a power-of-two tile width \( M \) known at compile time and \( 64 \)-byte-aligned allocation, this enables auto-vectorisation at the tile level.
  Cache-optimisation via tile blocking is well-established in the HPC literature~\cite{Park2003_TilingBlockDataLayoutAndMemoryHierarchyPerformance}.
  The closest prior art are the \emph{Block Hybrid} format~\cite{Gustavson2007_Algorithm865} and the SLATE dense linear algebra library~\cite{Gates2019_SLATE}, albeit tile sizes are optimised for Level-3 BLAS throughput rather than the power-of-two bit-flip access pattern of local quantum operation simulations.

  The total storage of \( N(N + M)/2 \) complex numbers exceeds the packed format by \( N(M-1)/2 \) (cf.~\Cref{tab:MemoryFootprint})---an overhead of less than \( 0.1\)\% for \( m = 5 \) at the benchmarking limit of \( n = 15 \) qubits (\Cref{sec:Results}).
  With double-precision complex arithmetic, a single tile occupies \( M^{2} \cdot 16\,\text{B} = 16\,\text{KiB} \) (for \( m = 5 \)), fitting comfortably in L1 cache (32--128\,KiB on modern CPUs) with some additional space for loop variables and register spilling. 
  Additionally, assigning tiles as atomic work units to the threads eliminates false sharing entirely, because distinct threads occupy disjoint sets of cache lines (\( 16 \)\,KiB \( \gg \) \( 64 \)\,B).

  \begin{table}[ht]
    \centering
    \begin{tabular}{rrrrr}
      \toprule
      \( n \) & Full & Packed & Tiled (\( m = 5 \)) & Tiled/Packed \\
      \midrule
      1  & 64\,B     & 48\,B     & 16\,KiB   & \( 341\times \) \\
      2  & 256\,B    & 160\,B    & 16\,KiB   & \( 102\times \) \\
      5  & 16\,KiB   & 8.25\,KiB & 16\,KiB   & \( 1.9\times \) \\
      10 & 16\,MiB   & 8.0\,MiB  & 8.25\,MiB & \( 1.03\times \) \\
      15 & 16\,GiB   & 8.0\,GiB  & 8.0\,GiB  & \( 1.001\times \) \\
      \bottomrule
    \end{tabular}
    \caption{%
      Memory footprint comparison (\( 16 \)\,bytes per complex double-precision number, \( m = 5 \), \( M = 32 \)).
      The Full column shows the \( N \times N \) hermitian matrix stored without exploiting symmetry.
      For \( n < m \), the tiled format allocates a single full tile (\( 16 \)\,KiB) regardless of the physical matrix size; the large overhead ratio is irrelevant in practice since the entire state fits in cache.}
    \label{tab:MemoryFootprint}
  \end{table}

  \subsection{Intra-tile and cross-tile operations}
  \label{sec:IntraCrossTile}

  \begin{algorithm}[t]
    \DontPrintSemicolon
    % Named data (typeset in sans-serif by algorithm2e)
    \SetKwData{Tile}{T}% % Pointer to first element of the tile
    \KwData{%
      Hermitian \( N \times N \) matrix \( H \) stored in \( \H[\ ] \) in tiled format with tile size \( M = 2^{m} \);
      complex \( 4 \times 4 \) transfer matrix \( S \); 
      target qubit \( a \in [0, m-1] \)%
    }
    \KwResult{Hermitian \( N \times N \) matrix \( H' = \Phi(H) \) stored in-place in \H[\ ]}

    \BlankLine
    \( n_{\mathrm{tiles}} \gets \lceil N/M \rceil \cdot ( \lceil N/M \rceil + 1) / 2 \) \tcp*[r]{Number of tiles}
    \( n_{\mathrm{base}} \gets \min(M, N) / 2 \) \tcp*[r]{Number of base indices}

    \BlankLine
    \tcc{Iterate lower-triangular tiles}
    \For{\( t \gets 0 \) \KwTo \( n_{\mathrm{tiles}} - 1 \)}{
      \( \Tile \gets \H + t \cdot M^{2} \) \tcp*[r]{First element in tile \( t \)}

      \BlankLine
      \tcc{Iterate row base-index pairs}
      \For{\( i \gets 0 \) \KwTo \( n_{\mathrm{base}} - 1 \)}{
        \( i_{0} \gets \)insertBits(\( i \), \( 0 \), \( \{a\} \)),\quad \( i_{1} \gets i_{0} + 2^{a} \)\;

        \BlankLine
        \tcc{Iterate column base-index pairs}
        \For{\( j \gets 0 \) \KwTo \( n_{\mathrm{base}} - 1 \)}{
          \( \ell_{00} \gets \)insertBits(\( j \), \( 0 \), \( \{a\} \))\(\,+\, i_{0} \cdot M \),\;
          \( \ell_{10} \gets \)insertBits(\( j \), \( 0 \), \( \{a\} \))\(\,+\, i_{1} \cdot M \),\;
          \( \ell_{01} \gets \ell_{00} + 2^{a} \),\quad\( \ell_{11} \gets \ell_{10} + 2^{a} \)\;

          \BlankLine
          \tcc{Gather block elements}
          \( v_{0} \gets \Tile[\ell_{00}],\quad
             v_{1} \gets \Tile[\ell_{01}] \),\;
          \( v_{2} \gets \Tile[\ell_{10}],\quad
             v_{3} \gets \Tile[\ell_{11}] \)\;

          \BlankLine
          \( w \gets S v \)\;

          \BlankLine
          \tcc{Update \( \H[\ ] \)}
          \( \Tile[\ell_{00}] \gets w_{0},\quad
             \Tile[\ell_{01}] \gets w_{1}\),\;
          \( \Tile[\ell_{10}] \gets w_{2},\quad
             \Tile[\ell_{11}] \gets w_{3} \)\;
        }
      }
    }
    \caption{%
      Update of the hermitian matrix, \( H \mapsto \Phi(H) \), stored in tiled format with \( M = 2^{m} \) according to a quantum operation acting on the single qubit \( a < m \).
      \label{alg:SingleQubitSingleTile}
    }
  \end{algorithm}

  The tile dimension \( m \) induces a natural partition of the update function.
  At the core of each update step lies the matrix vector multiplication \( w = S v \), but the central challenge is to systematically identify and access the correct entries of the hermitian matrix.
  While for \( k \)-local quantum operations acting on qubits \( A \) with \( a_{i} < m \) for \( i = 0, \dots, k-1 \) the update of the hermitian matrix involves only elements in a single tile, the number of tiles spanned by the local update for general \( A \) is proportional to \( 2^{2 \abs{A^{\ge m}}} \), where \( A^{\ge m} \coloneq \{ a_{i} \in A: a_{i} \ge m \} \).
  The corresponding branching, however, happens at the outermost level of the update function, introducing negligible overhead.

  For brevity, I will focus on single-qubit quantum operations presenting the two edge cases.
  The extensions to \( k = 2, 3 \) follow the same tile-level case structure with up to \( 2^{2k} \) tiles.

  \emph{Intra-tile} (\( a < m \)):
  A single-qubit quantum operation on qubit \( a \) couples elements residing in the same tile (cf. \Cref{alg:SingleQubitSingleTile}).
  The entire working set stays in L1 for the duration of the update, and the outermost loop over the \( N_{t}(N_{t}{+}1)/2 \) lower-triangular tiles is embarrassingly parallel.
  However, the non-contiguous access---the stride depends on \( a \)---impedes auto-vectorisation to the point of outweighing the L2 latency cost of the \emph{cross-tile} case.

  \emph{Cross-tile} (\( a \ge m \)): 
  A single-qubit operation on qubit \( a \ge m \) spans at most four tiles whose combined footprint (\( 4 \times 16\,\text{KiB} = 64\,\text{KiB} \)) fits in L2 cache, bounding performance by L2 bandwidth (cf. \Cref{alg:SingleQubitMultiTile}).
  Two loops, one nested inside the other, iterate through the subspace indices and the bit-insertion procedure~(\hyperref[sub:InsertBits]{Subroutine~\ref*{sub:InsertBits}}) consumes the relative target position \( a' = a-m \) within the tile index.
  Inserting bit values at position \( a' \) into the row and column tile indices yields three tiles that naturally belong to the stored lower triangle: \( T_{t_{i_{0}}, t_{j_{0}}} \) (bits \( 0,0 \)), \( T_{t_{i_{1}}, t_{j_{0}}} \) (bits \( 1,0 \)), and \( T_{t_{i_{1}}, t_{j_{1}}} \) (bits \( 1,1 \)).
  A three way branch at the tile level as opposed to the per-element branching of the packed format handles the tile \( T_{t_{i_{0}}, t_{j_{1}}} \).
  If \( t_{i} > t_{j} + 2^{a'} \), all four tiles can be accessed directly in the contiguous memory.
  Otherwise, if \( t_{i} < t_{j} + 2^{a'} \), then \( t_{j_{1}} > t_{i_{0}} \), so the tile \( T_{t_{i_{0}}, t_{j_{1}}} \) lies in the upper triangle and is accessed through its stored adjoint \( \adj{T}_{t_{j_{1}}, t_{i_{0}}} \), which lies in the lower triangle.
  Lastly, if \( t_{i} = t_{j} \), the tile \( T_{t_{i_{0}}, t_{j_{1}}} \) is in the upper triangle and its hermitian adjoint is precisely the tile \( T_{t_{i_{1}}, t_{j_{0}}} \).
  The subroutines, displayed in full detail in \hyperref[sec:Appendix]{Appendix~\ref{sec:AppendixA}}, differ only in their gather and scatter steps.
  Note that while subcase~C is triggered once for every tile row, subcase~A is the hot path for small \( a' \), whereas subcase~B dominates for large \( a' \), with a smooth transition between the two.

  \begin{algorithm}[t]
    \DontPrintSemicolon
    \KwData{%
      Hermitian \( N \times N \) matrix \( H \) stored in \( \H[\ ] \) in tiled format with tile size \( M = 2^{m} \);
      complex \( 4 \times 4 \) transfer matrix \( S \); 
      target qubit \( a \in [m, n-1] \)%
    }
    \KwResult{Hermitian \( N \times N \) matrix \( H' = \Phi(H) \) stored in-place in \H[\ ]}

    \BlankLine
    \( a' \gets a - m \) \tcp*[r]{Target tile index}
    \( n_{\mathrm{base}} \gets N / 2M \) \tcp*[r]{Number of tile base indices}

    \BlankLine
    \tcc{Iterate base-index of tile rows}
    \For{\( t_{i} \gets 0 \) \KwTo \( n_{\mathrm{base}} - 1 \)}{
      \( t_{i_{0}} \gets \)insertBits(\( t_{i} \), \( 0 \), \( \{a'\} \)),\quad
      \( t_{i_{1}} \gets t_{i_{0}} + 2^{a'} \)\;
    
      \BlankLine
      \tcc{Iterate base-index of tile columns}
      \For {\( t_{j} \gets 0 \) \KwTo \(  t_{i} \)}{
        \( t_{j_{0}} \gets \)insertBits(\( t_{j} \), \( 0 \), \( \{a'\} \)),\quad
        \( t_{j_{1}} \gets t_{j_{0}} + 2^{a'} \)\;

        \BlankLine
        \tcc{Tiles always in the lower triangle}
        \( \Tzz \gets \H + (t_{i_{0}} \cdot (t_{i_{0}} + 1) / 2 + t_{j_{0}}) \cdot M^{2} \)\;
        \( \Toz \gets \H + (t_{i_{1}} \cdot (t_{i_{1}} + 1) / 2 + t_{j_{0}}) \cdot M^{2} \)\;
        \( \Too \gets \H + (t_{i_{1}} \cdot (t_{i_{1}} + 1) / 2 + t_{j_{1}}) \cdot M^{2} \)\;

        \BlankLine
        \tcc{Subcase A: \( T_{t_{i_{0}}, t_{j_{1}}} \) in lower triangle}
        \uIf{\( t_{i_{0}} > t_{j_{1}} \)}{
          \( \Tzo \gets \H + (t_{i_{0}} \cdot (t_{i_{0}} + 1) / 2 + t_{j_{1}}) \cdot M^{2} \)\;
          crossTile(\( M \), \Tzz, \Tzo, \Toz, \Too, \( S \))\;
        }

        \BlankLine
        \tcc{Subcase B: \( T_{t_{i_{0}}, t_{j_{1}}} \) in upper triangle; read adjoint stored tile \( T_{t_{j_{1}}, t_{i_{0}}} \)}
        \uElseIf{\( t_{i} \neq t_{j} \)}{
          \( \Tzo \gets \H + (t_{j_{1}} \cdot (t_{j_{1}} + 1) / 2 + t_{i_{0}}) \cdot M^{2} \)\;
            crossTileAdj(\( M \), \Tzz, \Tzo, \Toz, \Too, \( S \)) \;
        }

          \BlankLine
          \tcc{Subcase C: \( t_{i} = t_{j} \); \Toz is the adjoint of \( T_{{i_{0}}, t_{j_{1}}} \)}
        \Else{
          crossTileDiag(\( M \), \Tzz, \Toz, \Too, \( S \)) \;
        }
      }
    }
    \caption{%
      Update of the hermitian matrix, \( H \mapsto \Phi(H) \), stored in tiled format with \( M = 2^{m} \) according to a quantum operation acting on the single qubit \( a \ge m \).
      \label{alg:SingleQubitMultiTile}
    }
  \end{algorithm}

  Assigning whole tiles as atomic work units to threads ensures that distinct threads touch disjoint cache lines, eliminating false sharing.
  Because all tiles have equal size, load balance is achieved by even distribution---unlike the packed format, where variable-length columns require round-robin assignment.

  The layout further benefits the computation of mean values
  \begin{equation*}
    \tr(\rho\,O) = \sum_{t_{i}} \operatorname{Re} \langle O_{t_{i}, t_{i}},\,\rho_{t_{i}, t_{i}} \rangle_{F} + 2 \sum_{t_{i} \ge t_{j}} \operatorname{Re} \langle O_{t_{i}, t_{j}},\,\rho_{t_{i}, t_{j}} \rangle_{F},
  \end{equation*}
  where both operators are stored in tiled format and \( \langle \cdot,\,\cdot \rangle_{F} \) is the \emph{Frobenius inner product}.
  Since each tile contribution is an independent scalar product on \( M^{2} \) contiguous elements, this is an embarrassingly parallel computation with each thread's working set fitting in L1 cache.

% ==========================================================================
\section{Results}
\label{sec:Results}
% ==========================================================================

% Present experimental setup, benchmarks, and results.

I now evaluate the practical performance of the newly introduced storage format against three established competitors, Qiskit Aer v0.17.2~\cite{JavadiAbhari2024_QuantumComputingWithQiskit, QiskitAer_v0.17.2}, QuEST v4.2.0~\cite{Jones2019_QuESTandHighPerformanceSimulationOfQuantumComputers, QuEST_v4.2.0}, and Qulacs v0.6.12~\cite{Suzuki2021_Qulacs, Qulacs_v0.6.12}, and one naive baseline.

\subsection{Benchmark setting}

  \begin{figure*}[t]
    \centering

    \begin{minipage}[b]{0.45\textwidth}
      \centering
      \includegraphics[width=\textwidth]{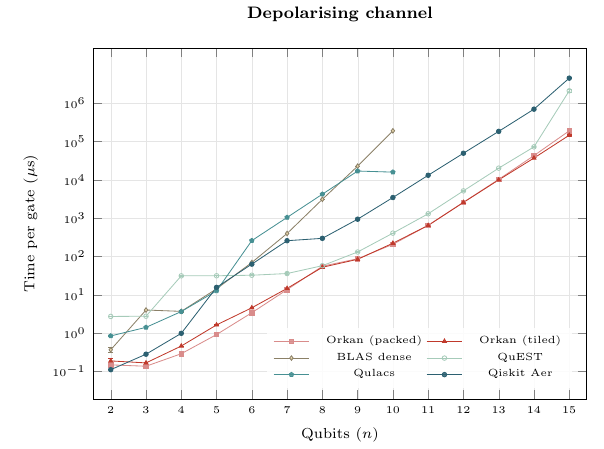}
      \phantomsubcaption\label{fig:BenchDepolarising-a}
    \end{minipage}%
    \hfill
    \begin{minipage}[b]{0.45\textwidth}
      \centering
      \includegraphics[width=\textwidth]{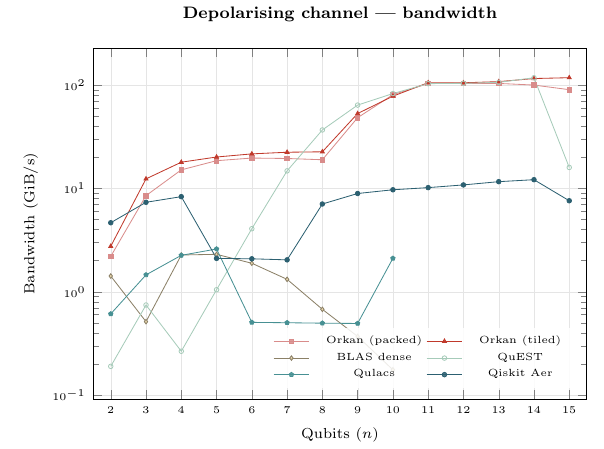}
      \phantomsubcaption\label{fig:BenchDepolarising-b}
    \end{minipage}
    \caption{
      (\subref{fig:BenchDepolarising-a})
        Time per execution of the single-qubit depolarising channel averaged over all possible qubit positions (\emph{lower is better}).
        The tiled and packed implementations separate from the field from \( n = 3 \) qubits on, with packed leading tiled until \( n \approx 12 \).
        The speedup settles to roughly \( 2 \times \) in the intermediate regime (\( 10 \le n \le 14 \)); at \( n = 15 \) the competitors enter an out-of-core regime in which the gap widens to \( 14 \times \).
      (\subref{fig:BenchDepolarising-b})
        Effective bandwidth of the single-qubit depolarising channel execution averaged over all possible qubit positions (\emph{higher is better}).
        The tiled and the packed implementations settle at \( \sim 20 \)\,GiB/s for \( n \le 8 \) qubits before increasing to \( \sim 120 \)\,GiB/s and \( \sim 100 \)\,GiB/s, respectively, for \( n > 8 \).
        For \( 11 \le n \le 14 \) this coincides with QuEST's throughput, suggesting the direct influence of the reduced memory footprint.
    }
    \label{fig:BenchDepolarising}
  \end{figure*}

  I compare two implementations shipped with the \emph{Orkan} library: the tiled format with \( m = 5 \) presented in \Cref{sec:Method} (hereafter \emph{tiled}) and the memory-optimal packed column-major format (hereafter \emph{packed}).%
  \footnote{Note that apart from the entries' ordering, i.e., row-major vs. column-major, one recovers the packed format from the tiled format by setting \( m = 0 \).}

  The baseline is a naive BLAS implementation of the Kraus map \eqref{eq:KrausRep}.
  The result of each matrix conjugation \( K_{\alpha} H\, \adj{K_{\alpha}} \) is added to the intermediate matrix, finally holding the updated hermitian matrix \( \sum_{\alpha} K_{\alpha} H\, \adj{K_{\alpha}} \), using BLAS matrix multiplication \texttt{zgemm}.
  The extended Kraus matrices \( K_{\alpha} = \id \otimes \cdots \otimes L_{\alpha} \otimes \cdots \otimes \id \) are assembled outside the timed region and accessed at runtime.

  The tests cover two complementary instances of a quantum operation.
  The single-qubit depolarising channel,
  \begin{equation*}
    \rho \mapsto (1 - p)\, \rho + \frac{p}{3}\,(X \rho X + Y \rho Y + Z \rho Z)
  \end{equation*}
  with fixed probability \( p = 0.1 \), triggers the most generic implementation of each backend and prevents them from using any gate-specific codepath.
  After receiving the corresponding local Kraus operators, i.e., \( \sqrt{1 - p}\, \id \), \( \sqrt{p / 3}\, X \), \( \sqrt{p / 3}\, Y \), \( \sqrt{p / 3}\, Z \), each backend converts them into their own native format outside of the timed region.
  The second instance is the Pauli-\( X \) gate as the prototype of native quantum gates---it avoids arithmetic operations altogether.
  While all benchmarked channels are trace-preserving, the algorithms apply unchanged to the trace non-increasing case.

  \begin{figure*}[t]

    \begin{minipage}[b]{0.45\textwidth}
      \centering
      \includegraphics[width=\textwidth]{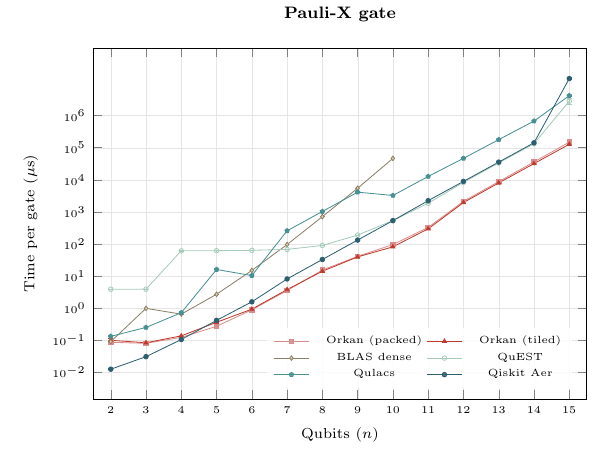}
      \phantomsubcaption\label{fig:BenchPauliX-a}
    \end{minipage}%
    \hfill
    \begin{minipage}[b]{0.45\textwidth}
      \centering
      \includegraphics[width=\textwidth]{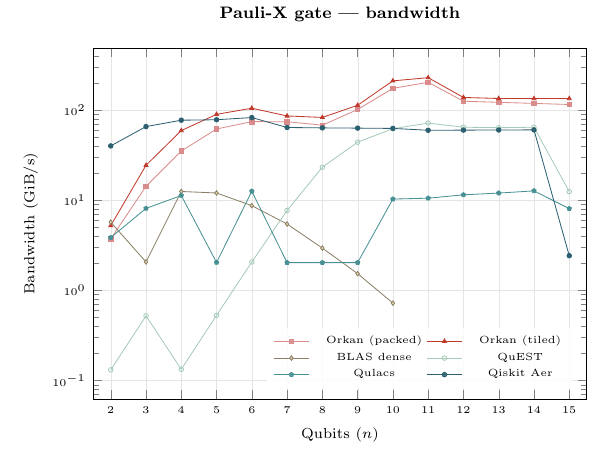}
      \phantomsubcaption\label{fig:BenchPauliX-b}
    \end{minipage}
    \caption{
      (\subref{fig:BenchPauliX-a})
        Time per execution of the Pauli-\( X \) gate averaged over all possible qubit positions.
        Except for small systems (\( n \le 4 \)), where Qiskit Aer's backend is slightly faster, the tiled and packed implementations exhibit a time saving by a factor of roughly four for intermediate systems (\( 10 \le n \le 14 \)).
        This additional factor of two is expected to stem from the double-pass for native gates in the established simulators used to profit from dedicated code paths.
      (\subref{fig:BenchPauliX-b})
        Effective bandwidth of the Pauli-\( X \) gate execution over all possible qubit positions.
        The tiled implementation quickly approaches \( \sim 100 \)\,GiB/s with a temporary jump to more than \( 200 \)\,GiB/s, consistently more than double the throughput of the reference backends.
    }
    \label{fig:BenchPauliX}
  \end{figure*}

  The analysis has two goals.
  First, I wish to determine whether storing only the independent entries of a hermitian matrix, beyond the obvious memory savings, also yields wall-clock speedups over established density-matrix simulations.
  Second, it aims to disentangle the contributions of reduced memory footprint and single-pass conjugation for native gates via an effective bandwidth analysis.

  I benchmark the wall time of a single channel execution averaged over all qubit positions.
  The hermitian operator is randomly initialised and, in general, neither positive semi-definite nor normalised.
  The benchmark circuit comprises \( L \) layers, each applying the channel sequentially to all \( n \) target qubits (single-qubit channel).
  To suppress statistical noise from clock imprecision, which is most pronounced at small \( n \), the layer count \( L \) is tuned per \( n \).
  Each test begins with a warm-up cycle to reduce initial cache latency.
  A single timing interval spans the full circuit and the input matrix is continuously evolved.
  I report \( 95 \)\% confidence intervals from \( 10 \) repetitions, even though in most cases they are too narrow to be distinguishable from the data marker itself.

  To evaluate whether the origin of any potential speedup is the reduced memory footprint of the new method, the effective bandwidth---a logical measure rather than actual DRAM traffic---of each process is captured as well.
  This means, for each method, the size of the stored data, i.e., \( N (N+M) / 2 \cdot 16 \)\,bytes for the tiled and packed implementations and \( N^{2} \cdot 16 \)\,bytes for the competitors and the baseline, is divided by their average execution time for an operation.

  All benchmarks were run on an Apple M3 Pro (6 performance + 5 efficiency cores) with \( 16.76 \)\,GiB of unified memory and a theoretical peak memory bandwidth of \( \sim 140 \)\,GiB/s, running macOS Sequoia 15.7.3.
  The code was compiled with Apple Clang 17.0.0 using \texttt{-O3 -march=native -fno-strict-aliasing} and OpenMP~\cite{Dagum1998_OpenMP} parallelisation with the maximum number of available threads.
  The \emph{Orkan} source code and the benchmark scripts used in this section are available at \url{https://github.com/Timo59/orkan} (tag \texttt{v0.1.0}).

\subsection{Benchmark results}

  \Cref{fig:BenchDepolarising} depicts the benchmark results for the depolarising channel.
  Both the packed and tiled implementations separate from the field at \( n = 3 \) and clearly dominate from \( n = 4 \) on, with the gap temporarily reaching \( 10 \)--\( 14 \times \) at \( n = 5, 6 \).
  For intermediate sizes (\( 10 \le n \le 14 \)), QuEST's validation overhead becomes negligible and the speedup settles to roughly \( 2 \times \), as the effective bandwidths (\Cref{fig:BenchDepolarising-b}) of the tiled implementation and QuEST converge to \( 100 \)\,GiB/s --- the memory-bandwidth limit where the amount of data to update becomes the decisive factor.
  At \( n = 15 \), however, QuEST suffers a sharp drop to below \( 20 \)\,GiB/s while both tiled and packed maintain their throughput, yielding a \( 14 \times \) speedup for the tiled implementation.
  Since the competitors' working sets at \( n = 15 \) consume \( 16 \)\,GiB, they barely fit into the available physical memory causing memory thrashing.
  This means that, due to inevitable background processes, the OS repeatedly swaps data between the RAM and the hard drive, while for the tiled memory, the \( \sim 8 \)\,GiB stays RAM-resident throughout.

  For small systems (\( n \le 7 \)), the packed format outperforms the tiled format by up to \( 80 \)\%.
  The logarithmic scale hides the fact that, after a transition phase \( 8 \le n \le 12 \) where runtimes nearly coincide, the tiled implementation pulls ahead with savings of \( 15 \)\% and \( 30 \)\% at \( n = 14 \) and \( n = 15 \), respectively.%
  \footnote{
    At small \( n \) the entire matrix already fits in L1/L2 regardless of layout, so the tiled format's \( N(M-1)/2 \) padding overhead dominates.
    The cache-locality benefit only manifests once the matrix exceeds the capacity of the lowest cache levels.
  }

  Qiskit Aer starts at just \( 0.1\, \mu s \) per channel application but quickly loses ground, settling at roughly \( 19 \times \) the runtime of the fastest method.
  The approach of manually iterating over four Kraus operators and summing intermediate results (BLAS, Qulacs) is uncompetitive for \( n > 7 \) due to repeated full-matrix passes, although the highly optimised Apple Accelerate BLAS backend remains competitive up to that point.
  These methods were capped at \( n = 10 \).

  The advantage of a dedicated treatment of hermitian operators becomes even more pronounced for native gate implementations and intermediate to large systems.
  \Cref{fig:BenchPauliX-a} shows that Qiskit Aer is the fastest backend for small system sizes (\( n \le 4 \)), owing to its dedicated Pauli code path.
  Nevertheless, the tiled implementation overtakes it at \( n = 6 \), revealing a speedup of \( 4 \times \) in the intermediate regime, before jumping to a \( 22 \times \) time saving at \( n = 15 \) over the best competitor, which is QuEST at that system size.

  The effective bandwidth (\Cref{fig:BenchPauliX-b}) supports the hypothesis that the asymptotic speedup is not solely attributable to the reduced memory footprint.
  After peaking at more than \( 200 \)\,GiB/s at \( n = 11 \)\footnote{At \( n = 11 \), the tiled working set of \( \approx 2 \)\,MiB still fits within the M3 Pro's system-level cache, so the figure reflects cache-resident rather than DRAM throughput.}, the asymptotic bandwidth of the tiled implementation settles at roughly twice that of Qiskit Aer and QuEST, suggesting that once the working set exceeds the last-level cache, decomposing the unitary conjugation into two separate passes doubles the memory traffic and halves the achievable throughput.
  Among the competitors, Qulacs is the only backend that uses the same generic gate path as for any other arbitrary single-qubit operation, with 16 complex multiplies per block instead of only memory swaps.
  This strategy proves to be inferior throughout, competing with the baseline for \( n \le 9 \) and settling with a slowdown of \( 5 \times \) over QuEST in the intermediate regime.

\medskip
In summary, the benchmark results confirm that the hermitian storage formats introduced in \cref{sec:Method} translate into consistent wall-clock speedups across all tested operations.
The advantage decomposes into roughly equal contributions from the reduced memory footprint and the single-pass conjugation.
At \( n = 15 \), the gap widens further as the competitors' \( 16 \)\,GiB working set triggers memory thrashing against the \( 18 \)\,GiB of physical memory; this effect should be verified on larger systems, since the competitors' density matrix grows to \( 64 \)\,GiB at \( n = 16 \).
The results further demonstrate that dedicated code paths for native gates are essential for competitive performance, as evidenced by the gap between Qulacs and QuEST.

% ==========================================================================
\section{Conclusion}
\label{sec:Conclusion}
% ==========================================================================

In this work, I presented \emph{Orkan}, a quantum computer simulator that specifically addresses the evolution of hermitian operators under the most general quantum operations.
By introducing the tiled memory layout, a layout well established in the HPC literature, to the field of quantum simulation, the memory footprint of a hermitian matrix is reduced by roughly \( 50 \)\% compared to the state-of-the-art methodology of storing the vectorisation of a density matrix.
Furthermore, I have provided numerical evidence that this memory format cuts the execution time of a generic quantum operation roughly in half and reduces the wall-clock time of native gate conjugations by a factor of approximately four, owing to the elimination of the two-pass overhead inherent in the vectorised approach.
Additionally, the tiled format mitigates the effects of false-sharing, per-element branching and load-imbalance of the packed column-major format at negligible storage overhead.
Beyond the speedup, the unified treatment of hermitian operators matches the operational setting of quantum computation, where every prediction is an expectation value and the choice between Schr\"{o}dinger and Heisenberg evolution is one of convention rather than physics.

Future work will extend the benchmark to multi-core x86 platforms and investigate distributed and GPU-accelerated variants, where the tile structure maps naturally onto shared-memory blocking.

% --------------------------------------------------------------------------
%  ACKNOWLEDGMENTS (optional)
% --------------------------------------------------------------------------
\section*{Acknowledgments}
I thank Lennart Binkowksi for launching the work on quantum simulators together with me and Ren\'{e} Schwonnek as well as Tobias Osborne for valuable input.
This work was supported by the DFG under Germany's Excellence Strategy--- EXC-2123 QuantumFrontiers, the Quantum Valley Lower Saxony and the BMFTR projects ATIQ and Quics.
The author used Claude Code for language editing and image generation throughout this article, for the implementation of the benchmark tests presented in \cref{sec:Results} and for distributing the code via GitHub.
All content was reviewed and edited by the author, who takes full responsibility for the final work.

% --------------------------------------------------------------------------
%  REFERENCES
%  Full papers: up to 2 extra pages for references.
%  Short papers: up to 1 extra page for references.
% --------------------------------------------------------------------------
\bibliographystyle{IEEEtran}
\bibliography{qsim, prelims, qcp}

\section*{Appendix}
\phantomsection\label{sec:Appendix}

\setlength{\floatsep}{2pt}
\setlength{\textfloatsep}{2pt}
\setlength{\intextsep}{2pt}
\SetAlCapSkip{2pt}

\subsection{Cross-tile subroutines}
\label{sec:AppendixA}

  \begin{procedure}[ht]
    \DontPrintSemicolon
    \KwData{
      Tile size \( M \);
      Tiles \( T_{t_{i_{0}}, t_{j_{0}}},\, T_{t_{i_{0}}, t_{j_{1}}},\, T_{t_{i_{1}}, t_{j_{0}}},\, T_{t_{i_{1}}, t_{j_{1}}} \) stored in row-major format to \Tzz, \Tzo, \Toz, \Too;
      complex \( 4 \times 4 \) transfer matrix \( S \)
    }
    \For{%
      \( k \gets 0 \) \KwTo \( M^{2} - 1 \)%
    }{

      \BlankLine
      \tcc{Gather block elements}
      \( v_{0} \gets \Tzz[k],\
         v_{1} \gets \Tzo[k],\
         v_{2} \gets \Toz[k],\
         v_{3} \gets \Too[k] \)\;

      \BlankLine
      \( w \gets S v \)\;

      \BlankLine
      \tcc{Update block elements}
      \( \Tzz[k] \gets w_{0},\
         \Tzo[k] \gets w_{1},\
         \Toz[k] \gets w_{2},\
         \Too[k] \gets w_{3} \)\;
    }
  \caption{crossTile($M$, $\protect\Tzz$, $\protect\Tzo$, $\protect\Toz$, $\protect\Too$, $S$)}
  \label{sub:crossTile}
  \end{procedure}

  This appendix details the three subroutines invoked by \Cref{alg:SingleQubitMultiTile}.
  Unlike the two-pass approach \( (\overline{U} \otimes \id)\,(\id \otimes U)\,\mathrm{vec}(\rho) \) used by established simulators (\Cref{sec:Background}), each subroutine applies the transfer matrix \( S \) in a single pass.
  All three subroutines implement the same gather--transform--scatter pattern: four coupled elements are read into a buffer, multiplied by \( S \), and written back.
  They differ only in how the fourth tile \( T_{t_{i_{0}}, t_{j_{1}}} \) is accessed, since it may lie outside the stored lower triangle.

  \hyperref[sub:crossTile]{Subroutine~\ref*{sub:crossTile}} handles the case \( t_{i_{0}} > t_{j_{1}} \), where all four tiles reside in the stored lower triangle.
  A single flat loop over the \( M^{2} \) element positions suffices because no per-element branching is required; this is the dominant code path for small \( a' \).

  \begin{procedure}[ht]
    \DontPrintSemicolon
    \KwData{
      Tile size \( M \);
      Tiles \( T_{t_{i_{0}}, t_{j_{0}}},\, T_{t_{i_{1}}, t_{j_{0}}},\, T_{t_{i_{1}}, t_{j_{1}}} \) stored in row-major format to \Tzz, \Toz, \Too;
      adjoint tile \( T_{t_{j_{1}}, t_{i_{0}}} \) to \Tzo;
      complex \( 4 \times 4 \) transfer matrix \( S \)
    }
    \tcc{\Tzo points to the adjoint tile: element \((i,j)\) of \( T_{t_{i_{0}}, t_{j_{1}}} \) is \( \overline{\Tzo[j,i]} \)}

    \BlankLine
    \For{\( i \gets 0 \) \KwTo \( M - 1 \)}{
      \For{\( j \gets 0 \) \KwTo \( M - 1 \)}{

        \BlankLine
        \tcc{Gather block elements}
        \( v_{0} \gets \Tzz[i \cdot M + j],\quad
           v_{1} \gets \overline{\Tzo[j \cdot M + i]} \),\;
        \( v_{2} \gets \Toz[i \cdot M + j],\quad
           v_{3} \gets \Too[i \cdot M + j] \)\;

        \BlankLine
        \( w \gets S v \)\;

        \BlankLine
        \tcc{Update block elements}
        \( \Tzz[i \cdot M + j] \gets w_{0},\quad
           \Tzo[j \cdot M + i] \gets \overline{w_{1}} \)\;
        \( \Toz[i \cdot M + j] \gets w_{2},\quad
           \Too[i \cdot M + j] \gets w_{3} \)\;
      }
    }
    \caption{crossTileAdj($M$, $\protect\Tzz$, $\protect\Tzo$, $\protect\Toz$, $\protect\Too$, $S$)}
    \label{sub:MultiTileAdj}
  \end{procedure}

  \hyperref[sub:MultiTileAdj]{Subroutine~\ref*{sub:MultiTileAdj}} handles \( t_{i_{0}} < t_{j_{1}} \) with \( t_{i} \neq t_{j} \), where the tile \( T_{t_{i_{0}}, t_{j_{1}}} \) lies in the upper triangle and is not stored explicitly.
  Its entries are recovered from the stored adjoint \( T_{t_{j_{1}}, t_{i_{0}}} \) by transposed, conjugated access on both read and write-back; the remaining three tiles are accessed in natural order.
  This branch dominates for large \( a' \).

  \begin{procedure}[ht]
    \DontPrintSemicolon
    \KwData{
      Tile size \( M \);
      Tiles \( T_{t_{i_{0}}, t_{j_{0}}},\, T_{t_{i_{1}}, t_{j_{0}}},\, T_{t_{i_{1}}, t_{j_{1}}} \) stored in row-major format to \Tzz, \Toz, \Too;
      complex \( 4 \times 4 \) transfer matrix \( S \)
    }
    \tcc{\Toz encodes both \( T_{t_{i_{1}}, t_{j_{0}}} \) and \( \adj{T}_{t_{i_{0}}, t_{j_{1}}} \)}
      \For{%
        \( i \gets 0 \) \KwTo \( M - 1 \)%
      }{
        \For{\( j \gets 0 \) \KwTo \( i \)}{

          \BlankLine
          \tcc{Gather block elements}
          \( v_{0} \gets \Tzz[i \cdot M + j],\quad
             v_{1} \gets \overline{\Toz[j \cdot M + i]} \)\;
          \( v_{2} \gets \Toz[i \cdot M + j],\quad
             v_{3} \gets \Too[i \cdot M + j] \)\;

          \BlankLine
          \( w \gets S v \)\;

          \BlankLine
          \tcc{Update block elements}
          \( \Tzz[i \cdot M + j] \gets w_{0},\quad
             \Tzz[j \cdot M + i] \gets \overline{w_{0}} \)\;
          \( \Toz[i \cdot M + j] \gets w_{2},\quad
             \Toz[j \cdot M + i] \gets \overline{w_{1}} \)\;
          \( \Too[i \cdot M + j] \gets w_{3},\quad
             \Too[j \cdot M + i] \gets \overline{w_{3}} \)\;
        }
      }
    \caption{crossTileDiag($M$, $\protect\Tzz$, $\protect\Toz$, $\protect\Too$, $S$)}
    \label{sub:Diagonal}
  \end{procedure}

  \begin{figure*}[t]

    \begin{minipage}[b]{0.45\textwidth}
      \centering
      \includegraphics[width=\textwidth]{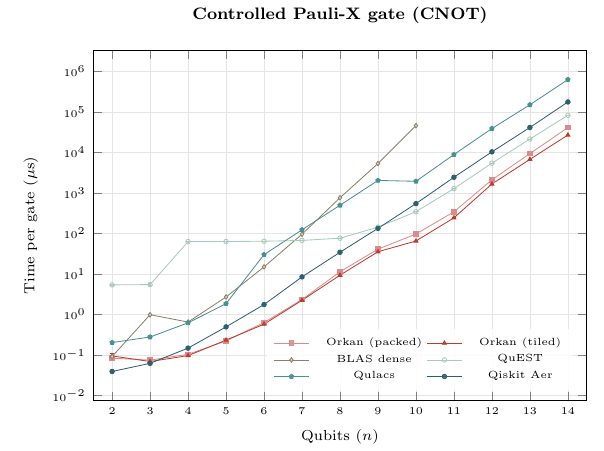}
      \phantomsubcaption\label{fig:BenchCnot-a}
    \end{minipage}
    \hfill
    \begin{minipage}[b]{0.45\textwidth}
      \centering
      \includegraphics[width=\textwidth]{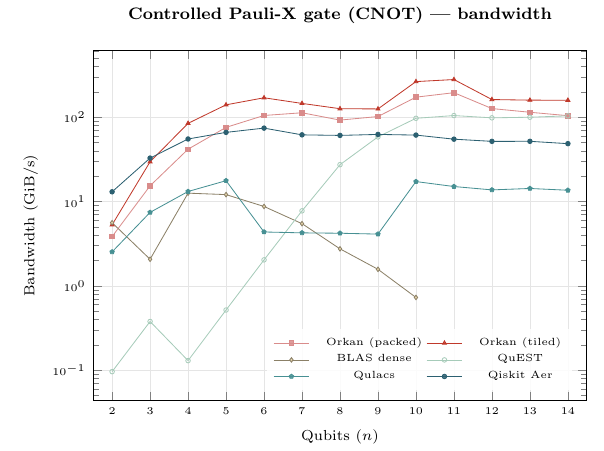}
      \phantomsubcaption\label{fig:BenchCnot-b}
    \end{minipage}
    \caption{
      (\subref{fig:BenchCnot-a})
        Time per execution of the controlled Pauli-\( X \) (CNOT) gate averaged over all possible qubit positions.
        In contrast to earlier native gates, the asymptotic speedup of the tiled and the packed implementation over QuEST is limited to a factor of three and two, respectively.
      (\subref{fig:BenchCnot-b})
        Effective bandwidth of the controlled Pauli-\( X \) (CNOT) gate execution over all possible qubit positions.
        Settling at \( \sim 150 \)\,GiB/s, the tiled implementation maintains an edge of more than \( 50 \)\% over the packed implementation as well as over QuEST.
        Remarkably, the tiled implementation achieves the highest effective bandwidth throughout all benchmark tests with \( \sim 300 \)\,GiB/s at \( n = 11 \).
    }
    \label{fig:BenchAppendix}
  \end{figure*}

  \hyperref[sub:Diagonal]{Subroutine~\ref*{sub:Diagonal}} handles the diagonal case \( t_{i} = t_{j} \), where the adjoint of \( T_{t_{i_{0}}, t_{j_{1}}} \) is precisely \( T_{t_{i_{1}}, t_{j_{0}}} \), so only three distinct tiles are involved.
  Moreover, the tiles \( T_{t_{i_{0}}, t_{j_{0}}} \) and \( T_{t_{i_{1}}, t_{j_{1}}} \) sit on the diagonal of the tile grid and are therefore hermitian; accordingly, the loop iterates only over the lower triangle (\( j \le i \)) within each tile and writes hermitian-symmetric counterparts explicitly.
  This branch is taken exactly once per outer-loop iteration of \Cref{alg:SingleQubitMultiTile}.

\subsection{More benchmark results}

  \begin{figure}[ht]
    \centering
    \includegraphics[width=\columnwidth]{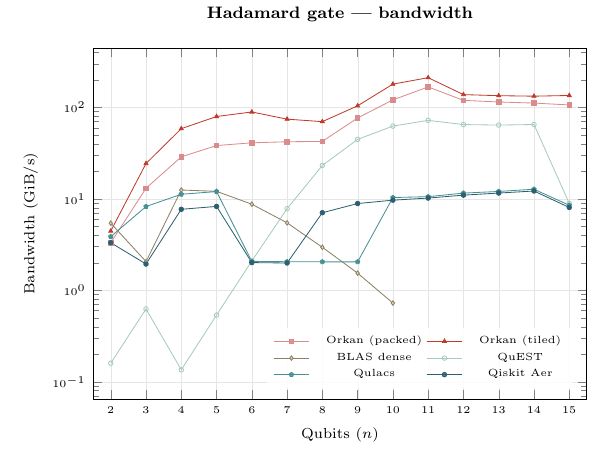}
    \caption{Effective bandwidth of the Hadamard gate execution over all possible qubit positions.
    After an intermediate peak of more than \( 200 \)\,GiB/s at \( n = 11 \), the tiled implementation converges to roughly \( 120 \)\,GiB/s, roughly doubling the throughput of QuEST.}
    \label{fig:BenchHadamard-b}
  \end{figure}

  \Cref{sec:Results} considered the depolarising channel and the Pauli-\( X \) gate as the two extreme cases in the benchmark set.
  The former is invoked by local updates via the transfer matrix, whereas the latter solely swaps entries in memory.
  In the following, I further corroborate the findings with the Hadamard gate---a non-zero arithmetic native gate---and the controlled Pauli-\( X \) (CNOT) gate---an even sparser permutation. 

  \Cref{fig:BenchHadamard-b} depicts the effective bandwidth of the Hadamard gate.
  While both, the tiled implementation and the packed implementation consistently outperform all competitors with an asymptotic speedup of \( 4 \times \), none of the competitors even constitute a considerable speedup over the baseline for \( n \le 7 \).
  This supports the observation that decomposing the unitary conjugation into two separate passes wastes roughly half of the achievable throughput, made in \Cref{sec:Results}.
  However, unlike in the former benchmark tests, the tiled implementation constantly outperforms the packed implementation from \( n = 6 \) on with a relative speedup growing up to roughly \( 25 \)\% for \( n = 15 \).

  \Cref{fig:BenchCnot-a,fig:BenchCnot-b} depict the benchmark results for the controlled Pauli-\( X \) (CNOT) gate.
  The contribution of the double-pass is less significant than for the other native gates as the asymptotic speedup of the tiled implementation over QuEST is reduced to a factor of three.
  However, this is expected because each pass of the state-vector kernel has to process only half of the elements---those with the control qubit set---whereas the single pass of \Cref{alg:k-localKraus} skips only one in four matrix elements---those with control qubit neither set in the row nor in the column index.
  This further reduced workload trivially leads to a higher effective bandwidth, reaching the maximum of \( 300 \)\,GiB/s over all benchmark tests at \( n = 11 \).
\end{document}